\documentclass{article}
\usepackage{amssymb}
\usepackage{amsmath} 
\usepackage{epsfig}
\begin{document}
\numberwithin{equation}{section}

\title{Raman solitons in transient SRS\footnote{to appear in Inverse Problems}} 
\author{M. Boiti$^+$, J-G. Caputo, J. Leon, F. Pempinelli$^+$ \\ 
{\em Physique Math\'ematique et Th\'eorique, CNRS-UMR5825,}\\ 
Universit\'e Montpellier 2, 34095 MONTPELLIER (France)\\
$(^+)$ {\em and Dipartimento di Fisica dell' Universit\`a}, Lecce (Italy)} 
\date{}\maketitle

\begin{abstract} We report the observation of Raman solitons on numerical
simulations of transient stimulated Raman scattering (TSRS)
with small group velocity dispersion.  The theory
proceeds with the inverse scattering transform (IST) for initial-boundary value
problems and it is shown that the explicit theoretical solution obtained
by IST for a semi-infinite medium fits strikingly well the numerical solution
for a finite medium. We understand this from the rapid decrease
of the medium dynamical variable (the potential of the scattering theory). 
The spectral transform reflection coefficient can be computed directly
from the values of the input and output fields and this allows to
see the generation of the Raman solitons from the numerical
solution. We confirm the presence of these nonlinear modes in
the medium dynamical variable by the use of a discrete spectral analysis.
\end{abstract}

\section{Introduction}\label{sec:basics}

Stimulated Raman scattering (SRS) is a 3-wave interaction  process with
extremely wide application in physics, especially in nonlinear optics
\cite{newell}\cite{yariv}. This essentially nonlinear phenomenon couples two
electromagnetic waves (pump and Stokes waves) to a two-level medium and is
described by a simple system of partial differential equations 
\cite{chusco}. 

Such a system applies to different physical situations, depending on a
phenomenological damping factor chosen to match the observed Raman linewidth. 
For instance, the steady state regime occurs when one neglects dynamical
effects on the medium. Then the system becomes explicitly solvable in terms of
the intensities and the result fits well the situation of strong damping and
long pulses, such as in fiber guides \cite{agrawall}. When instead dynamical
effects and damping are of same order, very interesting phase effects have been
discovered in \cite{druwen}. These have been interpreted as the manifestation
of solitons which, in the spectral transform scheme, would be related to 
discrete eigenvalues. Instead they are related to the continuous
spectrum  (they are not solitons) and named {\em Raman spikes} \cite{leon}. 
Therefore the question of the observation of the Raman soliton is open since the
original work \cite{chusco} in 1975.  

When the damping term is much smaller than
the dynamical response of the medium, and can be neglected
the regime is called  hyper-transient and applies to short duration pulses. 
Here the system possesses a Lax pair \cite{chusco,kaup,steudel} and can be
treated by means of  the inverse scattering transform (IST) generalized to
boundary-value problems \cite{jl,gakhov}.  The observation of the Raman soliton
is an interesting open problem, especially since the boundary-value problem for
TSRS has been completely solved on the finite interval in \cite{fok-men} with
the essential result that, as expected from the works
\cite{pavel,menyuk,winter}, the field $q(x,t)$ universally evolves toward the
self-similar solution.

The commonly used TSRS model is obtained as a 3-wave interaction
process where the group velocities of the 3 fields involved are considered
equal. As shown in \cite{sasha}, this assumption is an asymptotic
limit and consequently the group velocity
dispersion (GVD) has to be taken into account (through the spectral extension
of the input laser fields), leading to a modified SRS system (eq.  \eqref{srs}
below).  The main consequence of this fact is that Raman solitons (discrete
eigenvalues in the nonlinear Fourier spectrum) are effectively generated by SRS
in the medium \cite{sasha}.

The aim of this paper is to show that Raman solitons are indeed created in a
finite length medium and that their generation is accompanied with poles in the
nonlinear Fourier spectrum (spectral transform). We follow the model
derived and solved in \cite{sasha} and establish the following
results.

1 -  Based on the assumption of a  semi-infinite medium, IST gives an
explicit solution\footnote{The solution is not explicit for the finite length
case: both results of \cite{sasha} and of \cite{fok-men} give the answer by
solving a system of Cauchy-Green integral equations.} which will be shown to
be an excellent approximation of the finite medium case. We explain
this from the fast decay of the {\em potential}
$q(x,t)$ for large $x$.  Such is not the case in the zero-GVD case for which
$q(x,t)$ decreases as $x^{-3/4}$ typical of the self-similar solution
\cite{pavel,fok-men}.

2 - The reflection coefficient $\rho$ of the spectral transform 
(or nonlinear Fourier spectrum) is expressed in
terms of the input and output field envelopes, allowing us to check 
the appearance of Raman solitons (the real valued single poles of
$\rho$) on the numerical solutions. This also provides a means
to observe the generation of Raman solitons in experimental data.

3 - A recursion formula for computing the spectral transform of a finite set of
data has been recently proposed \cite{nft}.  We have implemented this
{\em discrete spectral transform} and confirmed the creation of these
Raman solitons in a simulation on a finite length.
A remarkable feature here
is that the numerical spectral transform can be quite easily implemented, 
its computation is faster, and it provides more information than a
conventional Fourier transform.

After presenting the model in section 2, we report the numerical 
observation of the Raman solitons in section 3. Section 4 explains
this by an analysis of the medium dynamical variable.

\section{The model}\label{sec:mod}

We present here the extension of the SRS model of \cite{chusco} to the case
where the dispersion of the group velocity is not neglected.
We consider the classical model as in \cite{yariv}, well adapted
to molecular Raman scattering, and for which the medium is schematically
represented by a collection $X$ of harmonic oscillators coupled to the electric
field $\vec E$ through the polarizability of the medium.

\subsection{Basic equations}\label{sub:base}

When the polarizability depends on the frequency of the applied field, the
group velocity of the electromagnetic waves becomes frequency dependent. While
a very small group velocity dispersion (GVD) has no consequence on the evolution
of the amplitude, it has nontrivial effects on the phase dynamics. In this
context, it has been proved in \cite{sasha}, using {\em multiscale
analysis}, that for the electric field 
\begin{align}\label{electric}
E(x,t')=& e^{i(k_1x-\omega_1t')}\int dk\ a (k,x,t)\ e^{-ikx}\notag\\
+& e^{i(k_2x-\omega_2 t')}\sqrt{\frac{\omega_2 }{\omega_1}}
\int dk\ b (k,x,t)\ e^{ikx}+ c.c.
\end{align}
and medium dynamical variable
$q(x,t) e^{i(Kx-\Omega t')} + c.c.$
where $\omega_1 -\omega_2 = \Omega$ and $k_1 - k_2= K$,
the resulting model of transient SRS can be written in the {\em retarded time} 
$t=t'-x/v$
\begin{align}\label{srs} 
&\partial_xa =qb e^{2ikx}\ ,\quad
\partial_xb =-\bar qa e^{-2ikx}\ , \notag\\ 
& \partial_tq =-g\int dk\ a \bar be^{-2ikx}\ , \end{align}
where the overbar stands everywhere for the complex conjugate.
Note the conservation ($x$-independence) of the flux
density $|a|^2+|b|^2$.  

The above model differs from the usual SRS system, corresponding
to the zero GVD case 
\begin{equation}\label{srs-sharp} 
\partial_xa  =q_0 b  \ ,\quad
\partial_xb  =-\bar q_0 a  \ , \quad
\partial_tq_0  =-g\, a  \bar b \ , \end{equation}
by the presence of the integral over all possible realizations of the phase 
mismatch $k$.

\subsection{Boundary value problem}\label{sub:bounds}

The question of interest is the time evolution of a couple $(a,b)$ of pump and 
Stokes
pulses of time duration $T$ sent into a medium  of length $\ell$. Hence the
domain of integration is
\begin{equation}\label{domain}
x\in[0,\ell]\ ,\quad t\in[0,T]\ .\end{equation}
and it is necessary to prescribe the values of the fields on the two boundaries
$t=0$ and $x=0$. The medium is initially ($t=0$) at rest (all molecules in the
fundamental state), hence the initial datum for the field $q$ is
\begin{equation}\label{init}
q(x,0)=0\ ,\end{equation}
independently of the reference frame 
\footnote{Note that a {\em prepared} medium would
correspond to a state $q$ given at {\em physical time} zero, that is on the
characteristic $t=-x/v$.}.

The input ($x=0$) light pulses are arbitrary functions of time $t$ with a given
spectral distribution around $k=0$, namely
\begin{equation}\label{input}
a(k,0,t)=A(k,t)\ ,\quad 
b(k,0,t)=B(k,t)\ .\end{equation}
We shall be working here with the representative example treated in
\cite{sasha} which consists in assuming a common spectral lineshape for $A$ and
$B$ as a normalized Lorentzian lineshape for the intensities. More precisely we
set
\begin{equation}\label{in-spectr}
|A(k,t)|^2=|A_0(t)|^2\ \frac1\pi\frac{\kappa}{k^2+\kappa^2}\ ,\quad
|B(k,t)|^2=|B_0(t)|^2\ \frac1\pi\frac{\kappa}{k^2+\kappa^2}\ ,
\end{equation}
where the input pulse profiles $A_0(t)$ and $B_0(t)$ are arbitrary. 

It is useful also to understand the scale invariance of the SRS
system, including its boundary values. By a simple  change of variables it can
be easily shown that the system  \eqref{srs} (characterized by the parameters
$g$ and $\ell$) {\em together with} the input boundary values 
\eqref{input} \eqref{in-spectr}
(characterized by the parameter $\kappa$) bears the  scale invariance
\begin{equation}\label{scale-param} 
\ell\to\alpha\ell\ ,\quad g\to g/\alpha\ ,\quad\kappa\to\kappa/\alpha\ .  
\end{equation} 
This invariance has been checked on numerical simulations as one way to test
the code accuracy, and it works so well that we have obtained indiscernible
pictures.

\section{Laser fields}

We describe hereafter the IST solution of the boundary-value problem 
\eqref{input} for the system \eqref{srs} using the results of \cite{sasha},
and compare it with the direct numerical solution in the finite length case.

\subsection{Sketch of the direct problem}

The direct problem consists in defining the spectral transform from the
data of the {\em potential} $q(x,t)$ and the boundary values $A(k,t)$
and $B(k,t)$ (as time $t$ appears everywhere as an external parameter,
we shall forget it here).
This is done by defining the following Jost solutions (for $k\in{\mathbb R}$)
\begin{align}\label{jost-sol}
&\varphi(k,x)=1-\int_0^xd\xi\ q(\xi)\phi(k,\xi)\ ,\notag\\
&\phi(k,x)=\int_0^xd\xi\ \overline{q}(\xi)\varphi(k,\xi)
\ e^{2ik(x-\xi)}\ ,
\end{align}
which are both entire functions of $k$ vanishing  as 
$k\to\infty$ in the upper half-plane.
\footnote{The relation with notations of \cite{sasha} is:
$\varphi(k)=\varphi_{11}^+(k)=\overline{\varphi_{22}^-}(\bar k)$,
$\phi(k)=-\varphi_{21}^+(k)=
\overline{\varphi_{12}^-}(\bar k).$}

These two functions then allow to define the {\em reflection coefficient}
$\rho(k)$ and the {\em transmission coefficient} $\tau(k)$ by taking the
limit $x\to\infty$ of $\varphi$ and $\overline{\phi} e^{2ikx}$ for
$k\in \mathbb R$, namely
\begin{equation}\label{scatt-coeffs}
\frac1{\tau}=1-\int_0^\infty d\xi\ q(\xi)\phi(k,\xi)\ ,\quad
\frac{\rho}{\tau}=-\int_0^\infty d\xi\ q(\xi)\overline{\varphi}(k,\xi)
e^{2ik\xi}\ .
\end{equation}
The coefficient $1/\tau$ is clearly an entire function of $k$ and one can show
\cite{sasha} that the reflection coefficient $\rho(k)$ is meromorphic in the 
upper half-plane with a finite number of single poles related to the
solitonic part of the solution $q$. We have also the following {\em unitarity} 
relation for $k\in{\mathbb R}$
\begin{equation}\label{unit}
|\rho|^2=1+|\tau|^2\ .\end{equation}

It is easy to prove finally that the vectors $(\varphi,\ -\phi e^{-2ikx})$ 
and $(\overline{\phi} e^{2ikx},\ \overline{\varphi})$ solve the same
differential equation as the vector $(a,\ b)$ in \eqref{srs}. Then by
comparing their values in $x=0$ we readily obtain from \eqref{input}
\begin{equation}\label{lin-rel}
a=A\varphi+B\overline{\phi} e^{2ikx}\ ,\quad
b=B\overline{\varphi}-A\phi e^{-2ikx}\ .\end{equation}

\subsection{Output pump pulse}

 From \eqref{scatt-coeffs}, \eqref{unit} and \eqref{lin-rel},
the output $|a(k,\ell,t)|^2$ is explicitly given for $\ell\to\infty$ by 
the expression 
\begin{equation}\label{pump-out-infty}
 |a (k,\infty,t)|^2=\frac1{1+|\rho|^2}\ 
\left|A-\rho B\right|^2\ ,\end{equation}
where $A$ and $B$  are the input data defined in \eqref{input}.

The main result of \cite{sasha} is that
the function $\rho(k,t)$ is obtained by solving the following Riccati time 
evolution
\begin{equation}\label{rho-evol}
\rho_t=-\rho^2\ {\cal C}_k^+[m^*]-2\rho\ {\cal C}_k^+[\phi]-
{\cal C}_k^+[m]\ ,\quad \rho(k,0)=0\ ,
\end{equation}
where the functions $m(k,t)$ and $\phi(k,t)$ are given from the input data by
\begin{equation}\label{m-phi}
m=\frac{i\pi}{2}g AB^*,\quad \phi=\frac{i\pi}{4}g (|A|^2-|B|^2)\ ,
\end{equation}
and where ${\cal C}_k^+$ denote the following Cauchy integral
\begin{equation}\label{cal-c}
{\cal C}_k^+[f]=\frac1\pi\int_{-\infty}^{+\infty}
\frac{d\zeta}{\zeta-(k+i0)}\ f(\zeta)\ .
\end{equation}
Consequently, for given inputs $A(k,t)$ and $B(k,t)$, the expression
\eqref{pump-out-infty} gives the explicit asymptotic output pump intensity
from the solution of the evolution \eqref{rho-evol}. 
It is worth remarking that, at any given time, $\rho$ possesses possibly
a finite number of simple poles whose time evolution is given by the 
nonlinearity of the Riccati equation.

For practical purpose, we choose in \eqref{input} the Stokes wave input seed as
a portion $e^{-\gamma}$ of the pump wave, namely
\begin{equation}\label{ratio}
B_0(t)=A_0(t)\  e^{-\gamma}\ .\end{equation}
In that case  the evolution \eqref{rho-evol} can be explicitly solved 
\cite{sasha} 
\begin{equation}\label{rho-sol-1}
\rho(k,t)=\frac{\sinh\delta(k,t)}{\cosh(\delta(k,t)-\gamma)},
\end{equation}
\begin{equation}\label{delta}
\delta(k,t)=\frac{i T(t)}{k+i\kappa},\quad
 T(t)=\frac14g(1+e^{-2\gamma})\int_0^td\tau|A_0(\tau)|^2.
\end{equation}

\subsection{Numerical solution of finite length TSRS}

Our purpose is to understand how the above IST-solution can be used to model
the solution of the SRS system \eqref{srs} on a finite length.  
This can be done first in a
qualitative way by writing the solution of \eqref{srs} with boundary values
\eqref{input} as the equivalent integral form
\begin{equation}\label{integral}
\left(\begin{array}{c} a(k,x,t) \\ b(k,x,t) \end{array}\right)=
\left(\begin{array}{c} A(k,t) \\ B(k,t) \end{array}\right)+
\int_0^xd\xi
\left(\begin{array}{c} q(\xi,t)b(k,\xi,t)e^{2ik\xi}\\ 
-\overline{q}(\xi,t)a(k,\xi,t)e^{-2ik\xi} \end{array}\right)\ .\end{equation}
The output $a(k,\ell,t)$ will not differ much from its asymptotic value
$a(k,\infty,t)$ if $q(x,t)$ is sufficiently small for $x>\ell$.  Consequently
the behavior of $q(x,t)$ at large $x$, is essential for the adequation of the
formula \eqref{pump-out-infty} to real situations.  We will see in the next
section that it is also crucial for the applicability of the spectral method. 
We now proceed to show that the numerical solution of the TSRS system
\eqref{srs} on a finite interval is in excellent agreement with the theoretical
expression for the output \eqref{pump-out-infty}.

We discretized \eqref{srs}
in both $x$ and $t$ using an order 2 Runge-Kutta method and advance
via the following algorithm:\\
1 - given $a(k,x,t)$ and $b(k,x,t)$ for all real k, compute $q(x,t)$ 
by integrating the time-evolution of $q$ at $x$ for the initial datum 
$q(x,t=0)=0$, where the integral is calculated using the trapezoidal rule,\\
2 - advance to $a(k,x+dx,t)$ and $b(k,x+dx,t)$, for all $k$,
by integrating the differential equation for $a$ and $b$,\\
3 - go to step 1 with $x=x+dx$.\\
The scheme is started at step 1 for $x=0$. 

The quality of the computation is monitored by evaluating the relative
error in the total flux $|a(k,x,t)|^2 + |b(k,x,t)|^2$ which is conserved
by \eqref{srs}. In all the runs that are presented it remained 
smaller than $10^{-5}$.

We chose as parameters
\begin{equation}\label{param}
\ell=80\ ,\quad g=0.5\ ,\quad \gamma=5\ ,\quad \kappa=0.2\ ,\end{equation}
and the input pump pulse envelope is the Gaussian
\begin{equation}
A_0(t)=\exp\left[-\left(\frac{t-50}{30}\right)^2\right]\ .
\end{equation}
For this choice we found that $dt \leq T/1000$ and $dx \leq \kappa / 2$
gave stable results. Another point is that because of the Lorentzian
line width we had to take a $k$ interval of width $ \approx 80 \kappa$
in order to ensure that the integrals were normalized. To describe
the strong oscillations present for the zero GVD system \eqref{srs-sharp}
we had to chose $dt \leq T/ 2000$ and $dx \leq \ell / 10000$.

A typical run with number of grid points in $x,t$ and $k$ $(n_x=n_t=1000;n_k=
500)$ takes about 2 hours CPU monoprocessor on a RS10000.  We used the
parallelism of the problem, i.e. the fact that the marching in $x$ (resp. $t$)
can be made in parallel for the loops in $t$ and $k$ (resp. $x$) and
implemented the code on a Silicon Graphics SGI 10000 using the OpenMP software.
This enabled a gain of a factor 8 or 10 in computing time depending on the
number of processors used.

\begin{figure}
\centerline{\psfig{file=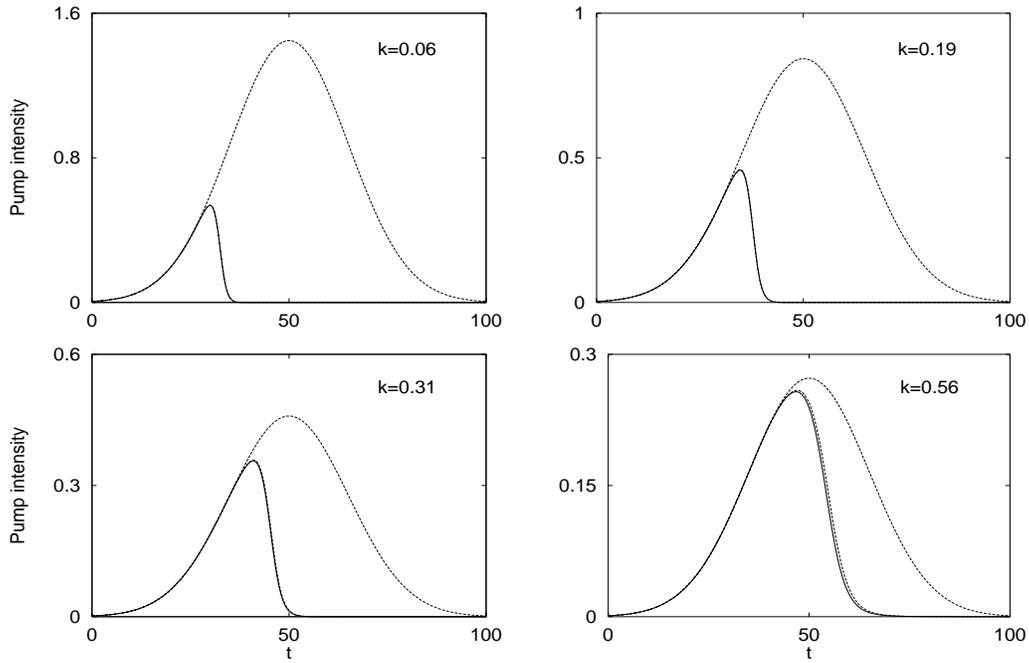,height=12cm,width=18cm}}
\caption{ Plot, as functions of time,  of the output pump intensity
$|a(x=\ell,k,t)|^2$ from the theoretical expression (\ref{pump-out-infty})
(full line) and the numerical solution of (\ref{srs}) (dashed line)
for four different values of $k$. 
The large gaussian curves (dashed line) represent the input pump intensities
$|a(x=0,k,t)|^2$.}\label{f01}
\end{figure}

Figure \ref{f01} show the pump intensity input $|a(k,0,t)|^2$ and output
$|a(k,\ell,t)|^2$ computed from the IST expression \eqref{pump-out-infty}
together with the numerical solution of the system \eqref{srs} for a length
$\ell=80$ and four different values of the parameter $k$, in excellent
agreement (a tiny discrepancy is only seen in the last picture).

It is worth remarking that one cannot pursue a given numerical experiment to a
longer length arbitrarily. We have stopped for instance the above calculation
at length $\ell=80$ where the potential $|q(x,t)| < 10^{-5}$. Continuing the
run to longer lengths would yield an increase of the potential which would then
start to oscillate.  This is probably due to the system itself which amplifies
the numerical errors in a drastic way at long lengths (note that from the scale
invariance \eqref{scale-param} longer length means larger Raman amplification).

\subsection{Raman solitons}

\begin{figure}
\centerline{\psfig{file=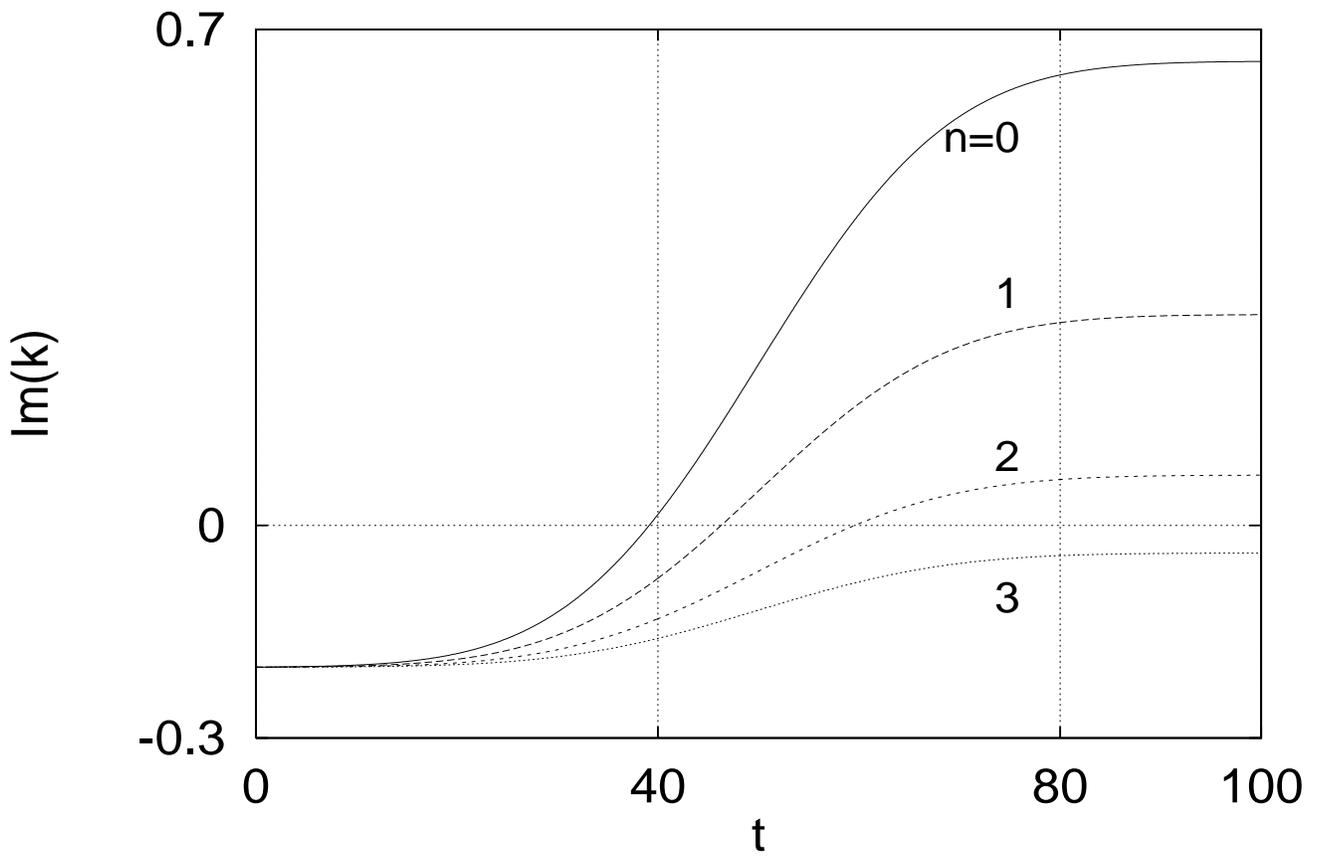,height=18cm,width=12cm,angle=-90}}
\caption{ Time evolution of the imaginary part of the poles $k_n$
of $\rho$ for $n=0,1,2$ and 3. The parameters are the same as in Figure 1}
\label{f02}
\end{figure}

The function $\rho$ of \eqref{rho-sol-1} has an essential singularity in 
$k=-i\kappa$ and a set of single poles $k_n$ evolving in time, given by
\begin{equation}\label{pole}
k_n(t)=-i\kappa+\frac{T(t)}{(n+\frac12)\pi-i\gamma}\ ,
\end{equation}
for $n\in{\mathbb N}$. At time zero no pole is present and, as $t$ evolves,
poles move upward from $-i\kappa$ and eventually reach the real axis at the
times $t_n$ defined by ${\rm Im}(k_n(t_n))=0$, i.e. by the implicit expression
\begin{equation}
T(t_n)=\frac{\kappa}{\gamma}\ [\gamma^2+[(n+\frac12)\pi]^2]\ .\end{equation}
Note that $t_n=t_{-n-1}$, hence the poles cross the real axis by pair. The
corresponding positions on the real axis are then given by
\begin{equation}\label{zetan}
k_n(t_n)=-k_{-n-1}(t_n)=\zeta_n\ ,\quad
\zeta_n=\frac\kappa\gamma(n+\frac12)\pi\ .\end{equation}
We plot in Figure \ref{f02} the imaginary part of the poles $k_n(t)$
as a function of time for the parameter values (\ref{param}) and
$n=0,1,2$ and 3. We observe that the first three poles cross the
real axis at the positions $\zeta_0=6.3 10^{-2}, \zeta_1=0.19, \zeta_2= 0.31$ 
and times $t_0=39.2, t_1=46.3$ and $t_2=59.6$. The pole $k_3$ starts
to evolve with $t$ but cannot cross the real axis because of the
finite duration of the pump pulse. 

As soon as these poles move to the upper half-plane, they generate a soliton
component in the {\em "potential"} $q(x,t)$. We will prove in the next section
that this potential (the medium dynamical variable) is a continuous 
function of $t$ when a pole crosses the real axis.

\subsection{The spectral transform from the output laser pulses}

The expressions \eqref{lin-rel} can be inverted to get the following
expressions of the Jost solutions in terms of the (physical) fields $a(k,x,t)$
and $b(k,x,t)$ and of the input pulses $A(k,t)$ and $B(k,t)$
\begin{equation}\label{inv-rel}
\varphi =\frac{a \overline{A}+\overline{b} B}{|A|^2+|B|^2} \ ,\quad
\overline{\phi}e^{2ikx}=\frac{a \overline{B}-\overline{b} A}{|A|^2+|B|^2} \ .
\end{equation}
Then  the definitions  \eqref{scatt-coeffs} 
lead to the following formula
\begin{equation}\label{rho-out}
\rho(k,t)=\left.
\frac{\overline{b} A-a \overline{B}}{a \overline{A}+\overline{b} B}
\right|_{x\to\infty}
\end{equation}
which gives the spectral transform $\rho(k,t)$ in terms of the output pump and
Stokes fields. This function for $k$ real must become singular at the two 
points $\pm \zeta_n$ 
each time a soliton $k_n$ is created and it is used now to prove 
the generation of Raman solitons in numerical experiments.

For the input pulses given in \eqref{in-spectr} and the particular choice
\eqref{ratio}, we readily obtain from the above
\begin{equation}\label{rho-out-th}
\rho=\left.\frac{\overline{b}-a e^{-\gamma}}{a +\overline{b} e^{-\gamma}}
\right|_{x\to\infty}\ .
\end{equation}
We use this expression to estimate $\rho_\ell$ from the numerical solution
$(a,b)$ at $x=\ell < +\infty$ and compare it to the $\rho$ obtained from the
IST (\ref{rho-sol-1}) for the parameters (\ref{param}).  Notice that
$|\rho(-k)|=|\rho(k)|$ so that we will only present positive values of $k$. 

\begin{figure} 
\centerline{\psfig{file=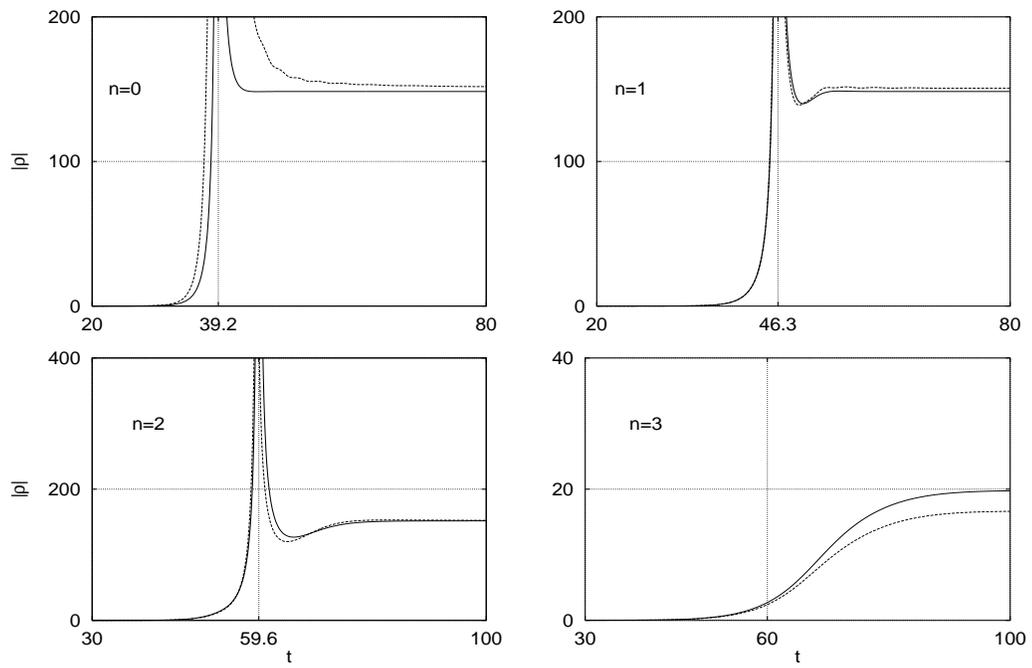,height=12cm,width=18cm,angle=0}}
\caption{Time evolution of the reflection coefficient $|\rho(\zeta_n)|$ 
for $n=0,1,2$ and 3 obtained from the inverse scattering
theory (\ref{rho-sol-1}) (full line) and from the numerical solution of 
(\ref{srs}) using
formula (\ref{rho-out-th}) (dashed line). }
\label{f03}
\end{figure}

Figure \ref{f03} shows the function $|\rho|$ of \eqref{rho-sol-1} in full line
and the numerical result obtained from \eqref{rho-out-th} in $x=\ell$ the in
dashed line, for the 3 values $k=\zeta_0,\zeta_1,\zeta_2$ and $\zeta_3$. Both
expressions are very close and as expected $\rho(\zeta_n)$ is singular for
$t=t_n$ for $n\leq 2$, while it is regular for $n=3$.  In the first picture of
figure \ref{f03}, the dashed line actually represents the value at $k=0$
instead of $k=\zeta_0$.  This is the only noticable discrepancy between
analytic asymptotic formula and numerical/experimental expression, resulting
from the finiteness in $x$ of the data $q(x)$ (the wave length $\lambda_0
\equiv 2 \pi / \zeta_0 \approx 104 > \ell$).

\section{Medium }

In the previous section we have seen that the finite length solution
is in good agreement with the asymptotic behavior 
(\ref{pump-out-infty}) given
by the IST on the semi-infinite line. We now proceed to justify this
fact by analyzing the behavior of the medium dynamical variable
$q(x,t)$.

\subsection{Sketch of the inverse problem}

IST furnishes the solution $q(x,t)$, called {\em medium dynamical variable},
by the expression \cite{sasha}
\begin{equation}\label{qrebuild}
q(x,t) =2i\psi^{(1)}(x,t)
\end{equation}
where $\psi^{(1)}$ is the coefficient of $1/k$ in the Laurent expansion
of the function $\psi(k,x,t)$ solution of
the following Cauchy-Green coupled system
\begin{align}\label{cauchy}
\varphi(k)=1+\frac1{2i\pi}\int_{-\infty}^{+\infty} d\lambda\ 
\frac{\bar\rho(\lambda)\psi(\lambda)}{\lambda-k-i0}\ e^{2i\lambda x}-
\sum_1^N\frac{\bar\rho(\bar k_n)\psi(\bar k_n)}{k-\bar k_n}\ e^{2i\bar k_n x}
\notag\\
\psi(k)=\frac1{2i\pi}\int_{-\infty}^{+\infty}
 d\lambda\ \frac{\rho(\lambda)\varphi(\lambda)}{\lambda-k+i0}
e^{-2i\lambda x}+
\sum_{n=1}^{N}\frac{\rho_n\varphi(k_n)}{k-k_n}\ e^{-2ik_nx} \ .
\end{align}
Here $\rho_n(t)$ are the $N$ residues of $\rho(k,t)$ at the poles $k_n(t)$ 
that are in the upper half $k$-plane (if any). 
Note that the solution $\varphi$ corresponds to the solution of 
\eqref{jost-sol} and that $\psi(k)=\overline{\phi}(\bar k)$.

Apparently the generation of a soliton is followed by the adjunction 
of a term in the equation for $\psi$, and consequently also in the 
expression of $q$, leading to a possible discontinuity for $q$.
We will now proceed to show that this is not the case.
This situation is particular to the spectral problem on the semi-line 
where the continuous spectrum (value of $\rho(k)$ on the real axis) 
is not separable from the discrete spectrum (poles of $\rho$ in the upper
half complex plane). This is due to the motion of the poles of $\rho$
that can cross the real axis. On the contrary, in the full line case, solitons
(discrete spectrum) can exist without radiation (continuous spectrum).

\subsection{Continuity}

 From now on when the spectral parameter $k$ is generically complex we shall be
using boldface letter ${\bf k}.$ From what precedes, at $t=t_{0}$ a pole
crosses the real axis at $k=k_0$ from the lower half plane. We consider here
the limits $t\rightarrow t_{0}\pm 0$ and we  write for $k\in{\mathbb R}$
\begin{equation}
\rho_{\pm }(k)= \rho (k,t_{0}\pm 0)=\frac{R_{0}(k)}{k-k_0\mp i0}
\label{rho+-}
\end{equation}
where $R_{0}({\bf k})$ is analytic in a neighborhood of $k_0$.
Note that 
\begin{equation} \label{rho+rho-}
\rho _{+}(k)=\rho_{-}(k)+2\pi i\rho_{0}\delta (k-k_0)
\end{equation}
where $\rho_{0}=R_{0}(k_0)$ is the residue of $\rho$ at the pole 
$k=k_0$. 

Assuming for simplicity no other pole, we have
from \eqref{cauchy} at $t=t_{0}-0$ 
\begin{equation}\label{cauch1}
\psi(k,x,t_{0}-0)=
{\frac{1}{2i\pi }}\int_{-\infty }^{+\infty }d\lambda 
\ \frac{\rho_{-}(\lambda)\varphi(\lambda,x,t_{0}-0)}{\lambda -k+i0}
\ e^{-2i\lambda x}
\end{equation}
which is well defined at $k=k_0$ since the distribution 
$(\lambda -k_0+i0)^{-2}$ is meaningful. 

On the contrary at $t=t_{0}+0$, expression \eqref{cauch1} diverges
due to the distribution $(\lambda -k_0+i0)^{-1}(\lambda -k_0-i0)^{-1}$. 
Since now there is a pole in the upper half complex plane, we should
take the limit ${\rm Im}({\bf k})\to 0$ in the complete expression in 
\eqref{cauchy}.  i.e.
\begin{align}\label{cauch2}
\psi({\bf k},x,t_{0}+0) =&
{\frac{1}{2i\pi }}\int_{-\infty }^{+\infty }d\lambda 
\ \frac{\rho_{+}(\lambda)\varphi(\lambda,x,t_{0}+0)}{\lambda-{\bf k}}
\ e^{-2i\lambda x}\notag\\
&+\frac{\rho _{0}\varphi(k_0,x,t_{0}+0)}{{\bf k}-k_0}e^{2ik_0x}\ ,
\end{align}
for ${\rm Im}({\bf k})<0$. We obtain 
\begin{equation}
\psi(k,t_{0}-0)=\psi(k,t_{0}+0).
\end{equation}
and consequently from (\ref{qrebuild}) 
\begin{equation}
q(t_{0}-0)=q(t_{0}+0).
\end{equation}

\subsection{Asymptotic behavior}

We consider here for simplicity the case when solitons are not yet present.
Then from \eqref{qrebuild} and \eqref{cauchy}, $q$ can be reconstructed in
terms of the reflection coefficient $\rho $ and the Jost solution $\varphi$ by
means of 
\begin{equation}\label{qrecons}
q(x)=-\frac{1}{\pi }\int_{-\infty}^{+\infty} 
dk\ \rho(k)\varphi(k,x)e^{-2ikx}\ .
\end{equation}
This formula yields directly an information on the behavior of $q$
at large $x$. 

In fact, if $\rho (k)\varphi(k,x),$ $\dots$, 
$\partial_{k}^{(n-1)}\{ \rho (k)\varphi(k,x)\}$ are
continuous and tend to $0$ for $k\to\infty $,  and if  
$\partial_{k}^{(n)}\{ \rho(k)\varphi(k,x)\} \in L({\mathbb R})$, we obtain by 
repeated integration by parts 
\begin{equation}
q(x)=\left( \frac{i}{2\pi x}\right)^{n}\int dk\ \partial _{k}^{(n)}\{
\rho (k)\varphi(k,x)\} e^{-2ikx}.
\end{equation}
Hence finally
\begin{equation}
x\to \infty\quad\Rightarrow\quad x^{n}q(x)\to 0\ .\end{equation}

As mentioned in the introduction, such an asymptotic behavior is not found for
the potential $q_0(x,t)$ which would be obtained from TSRS with zero GVD
\eqref{srs-sharp}.  Indeed, in that case the medium initially at rest evolves
universally towards the self-similar solution \cite{fok-men} which behaves as
$x^{-3/4}$ as found in \cite{pavel}. More precisely we have 
\begin{equation}\label{q-behav-sharp}
q\sim s(t)\,\xi^{-3/4}\left[\alpha\cos(h\xi^{1/2}+\beta)+{\cal O}(\xi^{-1/2})
\right]\ ,\quad \xi=s(t)\,x\ ,\end{equation}
where $s(t)$ is a function of $t$ determined by the boundary conditions and
where $\alpha$, $\beta$ and $h$ are arbitrary constants.

\begin{figure}
\centerline{\psfig{file=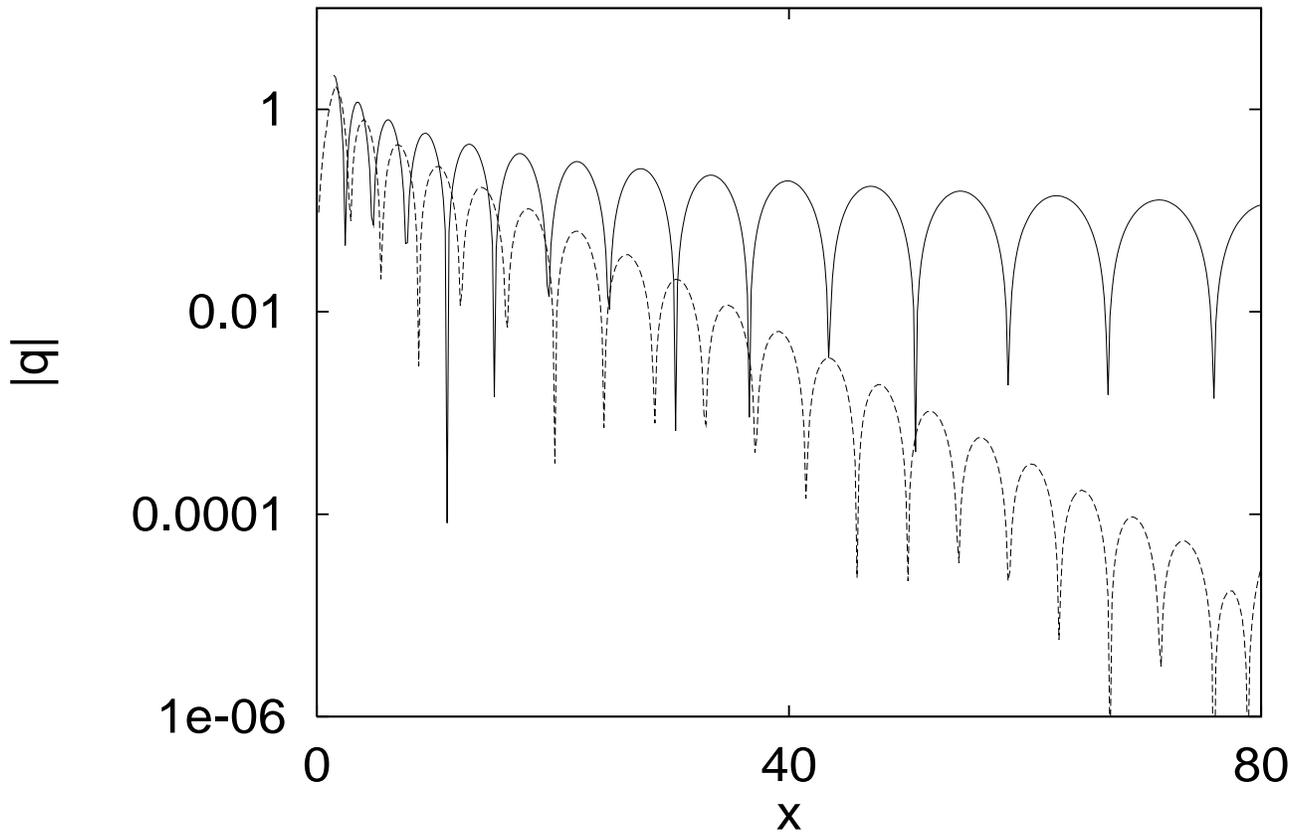,height=18cm,width=12cm,angle=-90}}
\caption{Evolution of $|q|(x,t=50)$ and $|q_0|(x,t=50)$ in linear-log
coordinates. }
\label{f04}
\end{figure}

Figure \ref{f04} presents the decay of $|q|$ and $|q_0|$ respectively in
dashed line and full line as a function of $x$ for $t=T/2=50$ in
a linear-log plot and the parameters \eqref{param}. The exponential
decay of $|q|$ can be seen for all values of $t$ and follows approximately
$e^{-\kappa x}$ as expected from the analysis. This observation justifies
the fact that the finite length evolution of \eqref{srs} is very close to
the asymptotic expression \eqref{pump-out-infty} given by IST. \\
On the contrary the slow decay of $|q_0|$ as $x^{-3/4}$ makes it impossible
for the integrals of \eqref{jost-sol} to exist so that the scattering theory must be 
revisited in this case.

\subsection{Numerical spectral transform and Raman solitons}

By numerical integration of the SRS system \eqref{srs} with a spatial grid of
dimension $h$, we get, at each value of time, a set of $L$ discrete data
$q(n,t)$ (with $x=nh$ and $\ell=Lh$), that we now analyze by means of the
spectral transform.  This is easily done, using the results of \cite{nft}
which, adapted to our notations, read
\begin{equation}\label{disr}
\rho(k,t)=R_0(\zeta,t)\ ,\end{equation}
where $R_0$ is obtained from the following inverse recursion
\begin{equation}R_L =\zeta^Lhq(L,t)  \ ,\quad
R_{m-1}=\frac{R_m-hq(m{-}1,t)\,\zeta^{m-1}}
{1+R_mh\bar q(m{-}1,t)\,\zeta^{-m+1}}\ ,
\end{equation}
and where the parameter $\zeta$ is related to $k$ by
\begin{equation}
\zeta=e^{2ikh}\ .\end{equation} 

\begin{figure}
\centerline{\psfig{file=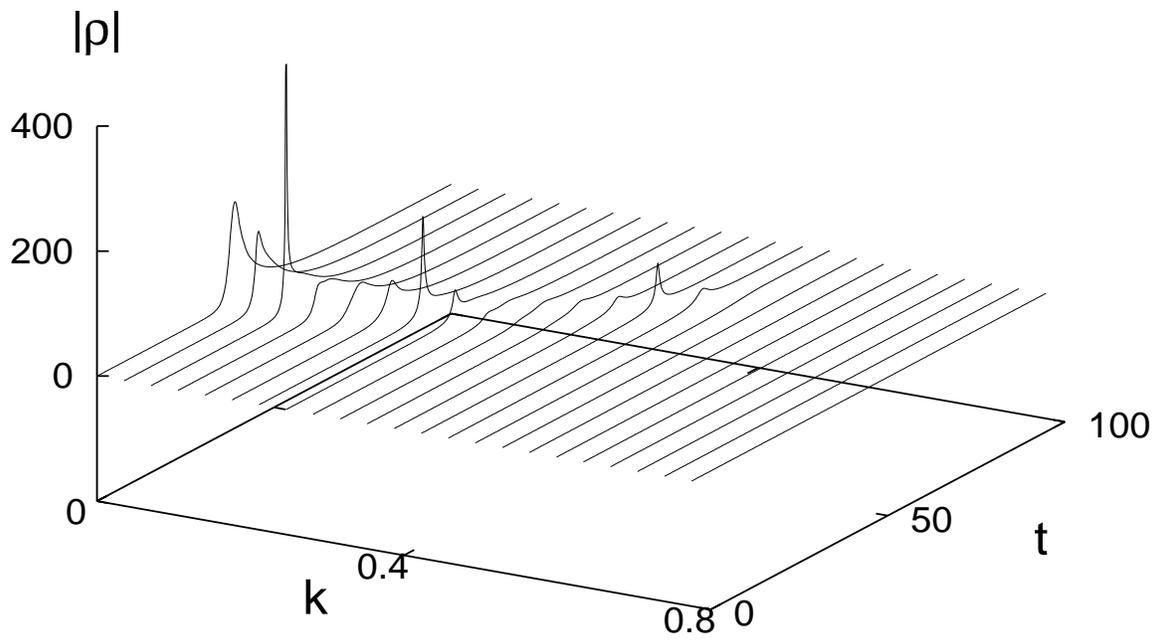,height=18cm,width=12cm,angle=-90}}
\caption{Evolution of $|\rho|(k,t)$ obtained from the
discrete spectral transform (\ref{disr}). }
\label{f05}
\end{figure}

It is then a simple task to use this recursion relation to compute the
numerical spectral transform $R_0(\zeta,t)$ and compare it to the asymptotic
theoretical expression \eqref{rho-sol-1} of $\rho(k,t)$. In Figure
\ref{f05} we plot $|\rho(k,t)|$ where one can clearly see the three
singularities $(\zeta_n,t_n)$ for $n=0,1,2$ corresponding to the three
poles of $\rho$.

The non linear spectral transform is easier to compute, faster and
yields more information than the standard Fourier transform which in
particular does not present any singularity (as we checked numerically).

\section{Conclusion}

This work demonstrates that the IST solution of transient SRS on the
semi-infinite line furnishes a very accurate model for the solution on a finite
domain. The accuracy stands not only for the output laser pulses intensities
but also for their phases as shown in a spectacular way by the generation of
the Raman solitons. It is shown that the question of the experimental
observation of Raman solitons is solved by a convenient combination of the
output (and input) laser profiles.  We also performed a numerical {\em
nonlinear spectral analysis} which resulted to be not only rich of information
but also quite easy to implement and faster than the usual FFT procedure.  

Finally we mention that it is difficult to compare these results with those
obtained in the zero GVD case \eqref{srs-sharp}. The problem is that in the
limit $\kappa\to0$, the spectral transform gets an essential singularity in
$k=-i0$ which sends poles to the upper half-plane. To discuss the number and
time location of such poles would require not only a careful study of the
singular limit $\kappa\to0$, but also to reformulate IST on the half-line as we
have seen that the potential $q(x)$ does not decrease {\em fast enough} as
$x\to\infty$.

\section*{Acknowledgments}
JGC is on leave from the Laboratoire de Math\'ematiques de
l'INSA de Rouen. The authors thank the CRIHAN computing
center for the use of their facilities. We thank INTAS
for support through grant 96-339 and support from Sezione INFN di Lecce 
and PRIN 97 "Sintesi".

\end{document}